\documentclass{ametsocV5}

\usepackage{amsmath,amsfonts,amssymb,bm}
\usepackage{mathptmx}
\usepackage{newtxtext}
\usepackage{newtxmath}
\nolinenumbers

\title{Detecting cold pools from soundings during EUREC$^4$A}

\authors{Ludovic Touzé-Peiffer\correspondingauthor{Ludovic Touzé-Peiffer, ludovic.touze-peiffer@lmd.ipsl.fr}, Raphaela Vogel}

\affiliation{LMD/IPSL, CNRS, Sorbonne University, Paris, France.}

\extraauthor{Nicolas Rochetin}

\extraaffil{LMD/IPSL, CNRS, ENS, Paris, France.}

\abstract{This paper develops a novel method to detect cold pools from atmospheric soundings over tropical oceans and applies it to sounding data from the EUREC$^4$A field campaign, which took place south and east of Barbados in January-February 2020. The proposed method exploits the fact that the air in a cold pool is denser than the air above it. It leads us to define cold pool soundings as those for which the mixed-layer height is smaller than 400 m. We first test this criterion by verifying its consistency with surface temperature and precipitation in a realistic high-resolution simulation over the western tropical Atlantic. Applying it to EUREC$^4$A data, we then identify 7 \% of EUREC$^4$A dropsondes and radiosondes as cold pool soundings. In two selected case studies, we find that cold pools soundings coincide with mesoscale cloud arcs and temperature drops in the surface time series. Statistics for the entire campaign further characterize the signature of cold pools in temperature, humidity and wind profiles. In the presence of wind shear, we show in particular that the spreading of cold pools is favored downshear, suggesting downward momentum transport by unsaturated downdrafts. These results support the robustness of our simple method in different environmental conditions and illustrate the new insights it offers for the characterization of cold pools and their environment.}

\begin{document}

\maketitle


\section{Introduction}

Below clouds, the partial evaporation of precipitation may cool the air sufficiently to generate unsaturated downdrafts, which spread horizontally when reaching the surface under the form of density currents. As these so-called "cold pools" expand, they lift the warmer adjacent air masses, leading in some situations to the creation of new convective cells \citep[e.g.][]{Craig1976, Warner1979}. This triggering has been observed to be particularly effective when two cold pools collide \citep{Meyer2020,Droegemeier1985, Feng2015} or when the vorticity created by a cold pool counteracts that from the low-level wind shear \citep{Rotunno1988, Weisman2004}. In addition, since they induce strong gusts near the surface, cold pools are suspected to enhance surface fluxes, and thus to modify the  properties of the subcloud layer (SCL) \citep{Tompkins2001, Langhans2015}. For all these reasons, cold pools are thought to play an important role for the organization and the propagation of convection \citep{Schlemmer2014, Tompkins2001, Kurowski2018}.

Cold pools are ubiquitous in regions of precipitating convection, both over lands and oceans \citep{Zuidema2017, Bryan2010}. Most past attempts to detect cold pools in observational data used methods based on surface time series. Analyzing data from the Tropical Ocean Global Atmosphere (TOGA) Coupled Ocean-Atmosphere Response Experiment \citep[COARE;][]{Webster1992}, \citet{Young1995} for instance defined the beginning of a cold pool time period through the onset of any rain shaft of at least 2 mm h$^{-1}$ and the end of a cold pool time period by the end of the subsequent surface temperature recovery. During the Rain in Cumulus over the Ocean campaign \cite[RICO;][]{Rauber2007}, which took place in the eastern Caribbean from December 2004 to January 2005, \citet{Zuidema2012} used a similar method but did not impose any threshold on surface rain rate, in order to consider all precipitating events. More recently, \citet{Szoeke2017} detected cold pools over the central Indian Ocean during the Dynamics of the Madden-Julian Oscillation experiment \citep[DYNAMO;][]{Yoneyama2013} as temperature drops in the surface time series measured by the research vessel Roger Revelle. A slightly modified version of this method was used  in \citet{Vogel2017} to detect cold pools from 2011 to 2017 at the Barbados Cloud Observatory (BCO), a site exposed to relatively undisturbed trade-winds on the eastern side of Barbados \citep{Stevens2016, Medeiros2016}.

Surface measurements from research vessels or weather stations are useful to get data with high temporal resolution at a single point. They give precise information on the surface characteristics of cold pools, such as the wind shifts they induce, the amplitude of the associated temperature drops, their effects on surface fluxes, etc. However, they say little about their vertical structure, in particular their height, which is recognized as an important parameter to describe the triggering of new convective cells near cold pool edges \citep{Grandpeix2010, Jeevanjee2015}. The distribution of temperature and moisture in and above cold pools is also poorly known \citep{Zuidema2017}, as is the distribution of horizontal and vertical winds. In a study of shallow marine cumulus convection, using large-scale simulations, \citet{Bretherton2017} found mesoscale convergence under precipitating cloud clusters and associated circulations that could explain the mesoscale aggregation of shallow convection under these conditions. Detecting similar circulations above cold pools to test this hypothesis in observations remains a challenge.

The field study EUREC$^4$A \citep[Elucidating the role of clouds-circulation coupling in climate;][]{Bony2017} might help overcome some of these difficulties. EUREC$^4$A was held in January and February 2020 over the Atlantic ocean east and south of Barbados \citep{Stevens2020overview} and involved a wide variety of platforms: among others, four research aircraft, four research vessels, surface observatories and a battery of uncrewed aerial and seagoing systems. Aircraft and research vessels supported the implementation of a large-scale sounding array. In total, more than 2000 atmospheric profiles were measured during EUREC$^4$A using radiosondes and dropsondes. This unprecedented dataset represents a unique opportunity to study the properties of cold pools over tropical oceans, provided a robust method to detect cold pools from atmospheric soundings exists. 

The lack of such a method motivates the present study, which proposes a detection method of cold pools based on the height of the mixed layer and applies it to EUREC$^4$A data. The paper is structured as follows: first, in section \ref{data}, we describe the EUREC$^4$A radiosonde and dropsonde data used in this study, then, in section \ref{method}, we present our detection method of cold pools and test it in a high resolution simulation. Finally, we assess our detection method with EUREC$^4$A data and use it to provide a first analysis of cold pools during the field study. 

\section{Radiosonde and dropsonde data}

\label{data}

During EUREC$^4$A, atmospheric soundings were released in two different regimes: the 'Tradewind Alley' and the 'Boulevard des Tourbillons' \citep{Stevens2020overview}. The Tradewind Alley is a corridor up to 50$^\circ$W east of Barbados, extending from approximately 11$^\circ$N to 17$^\circ$N. In January and February, free-tropospheric subsidence prevails in these latitudes and the Tradewind Alley can thus be seen as a trade-wind regime, with shallow convective clouds under a capping inversion \citep{Stevens2016}. The Boulevard des Tourbillons is a corridor further south extending from the northern coasts of Brazil to Barbados. It was initially chosen to characterize large-scale oceanic eddies observed along the coast of South America and their impact on air-sea interactions \citep{Bony2017, Stevens2020overview}. Located in areas of deeper convection, the Boulevard des Tourbillons is also an opportunity to extend the EUREC$^4$A sounding array down to the Intertropical Convergence Zone.

In the Tradewind Alley, both radiosondes and dropsondes were used to characterize the trade-wind environment. Radiosondes were launched from land at BCO (13.16$^\circ$N, 59.43$^\circ$W), and from two research vessels -- the German research vessel Meteor, cruising between 12-14.5$^\circ$N at a fixed longitude of 57$^\circ$W, and the American research vessel from NOAA Ronald H. Brown (RH-Brown), moving along transects further to the east. Two research aircraft, the German HALO and the WP-3D Orion (P-3) from NOAA, complemented this set of measurements by launching dropsondes. HALO flew at an altitude of about 10 km along the 'EUREC$^4$A circle', a circular path with approximately 200 km diameter centered at 13.3$^\circ$N, 57.7$^\circ$W. When launching dropsondes, the P-3 flew at about 7 km along both linear and circular patterns around the EUREC$^4$A circle, as well as further to the east close to the nominal position of the RH-Brown. No dropsondes were launched in the Boulevard des Tourbillons, but radiosondes were launched from the French research vessel L'Atalante and from the German research vessel Maria S. Merian (MS-Merian).

Equipped with a GPS receiver, all these radiosondes and dropsondes measured the pressure, temperature, humidity and horizontal wind along their fall or ascent. For radiosondes equipped with parachutes, we count the ascent and the descent as two separate profiles. In this study, we use quality controlled data (Level-3) interpolated on a common altitude grid with bin sizes of \text{10 m} \citep{Stephan2020, George2020}. In addition to the filters already used in the initial dataset, we are adding a filter to keep only soundings with at least 30 measurements of temperature, pressure and humidity below 500 meters. This filter is needed to get enough data in the lowest atmospheric layers to distinguish cold pools from their environment. It mainly affects descending profiles from radiosondes as their signal was sometimes lost in the first hundreds of meters above sea level due to Earth's curvature \citep{Stephan2020}. It removes in total 441 radiosonde profiles, resulting in an input dataset set with 1068 atmospheric profiles from dropsondes and 1106 from radiosondes. 

\section{Presentation of the method and test in a high-resolution simulation}

\label{method}

Most methods which have been proposed to detect cold pools from observations used the fact that the air inside a cold pool is colder than the air \textit{around} it. The present detection method uses the fact that cold pool air is colder than the air \textit{above} it. In terms of virtual potential temperature, a cold pool is indeed colder, thus denser than the SCL air on top of it. Consequently, a sharp increase in virtual potential temperature $\theta_v$ is expected at the top of a cold pool. Over tropical oceans, in convective regimes, it contrasts with the SCL outside cold pools, which tends to be well-mixed in $\theta_v$ up to cloud base \citep[e.g.][]{Pennell1974, Cuijpers1993}. This motivates the use of a cold pool detection method based on the height of the mixed layer ($H_{mix}$).

Following \citet{Canut2012} and \citet{Rochetin2020}, we define $H_{mix}$ as the lowest altitude $Z$ above $Z_{min}=100 \textrm{ m}$ where the virtual potential temperature $\theta_v$ is higher than its mass-weighted average from $Z_{min}$ to $Z$ by a fixed threshold $\epsilon = 0.2$ K:
\begin{equation}
  \begin{aligned}
   \theta_v(Z) &\geq \tilde{\theta_v} + \epsilon\\
  \textrm{with } \tilde{\theta_v} &= \frac{\int_{Z_{min}}^{Z} \rho(z) \theta_v(z) dz}{\int_{Z_{min}}^{Z} \rho(z) dz}
  \end{aligned}
\end{equation}
$\rho$ being the density of the air. Setting $Z_{min}$ at 100 m is necessary due to the presence of unphysical temperature peaks below 100 m for a few radiosondes. The virtual potential temperature is calculated assuming that the air of the lowest layers is not saturated, so that the mixing ratio of liquid water in the air can be neglected. It is then approached as: $\theta_v = \theta (1+0.61r)$, r being the mixing ratio of water vapor. The calculation of $H_{mix}$ thus requires only the vertical profiles of pressure, temperature and humidity at a single point and is directly applicable to soundings.

\begin{figure}[t]
        \noindent\includegraphics[width=\textwidth]{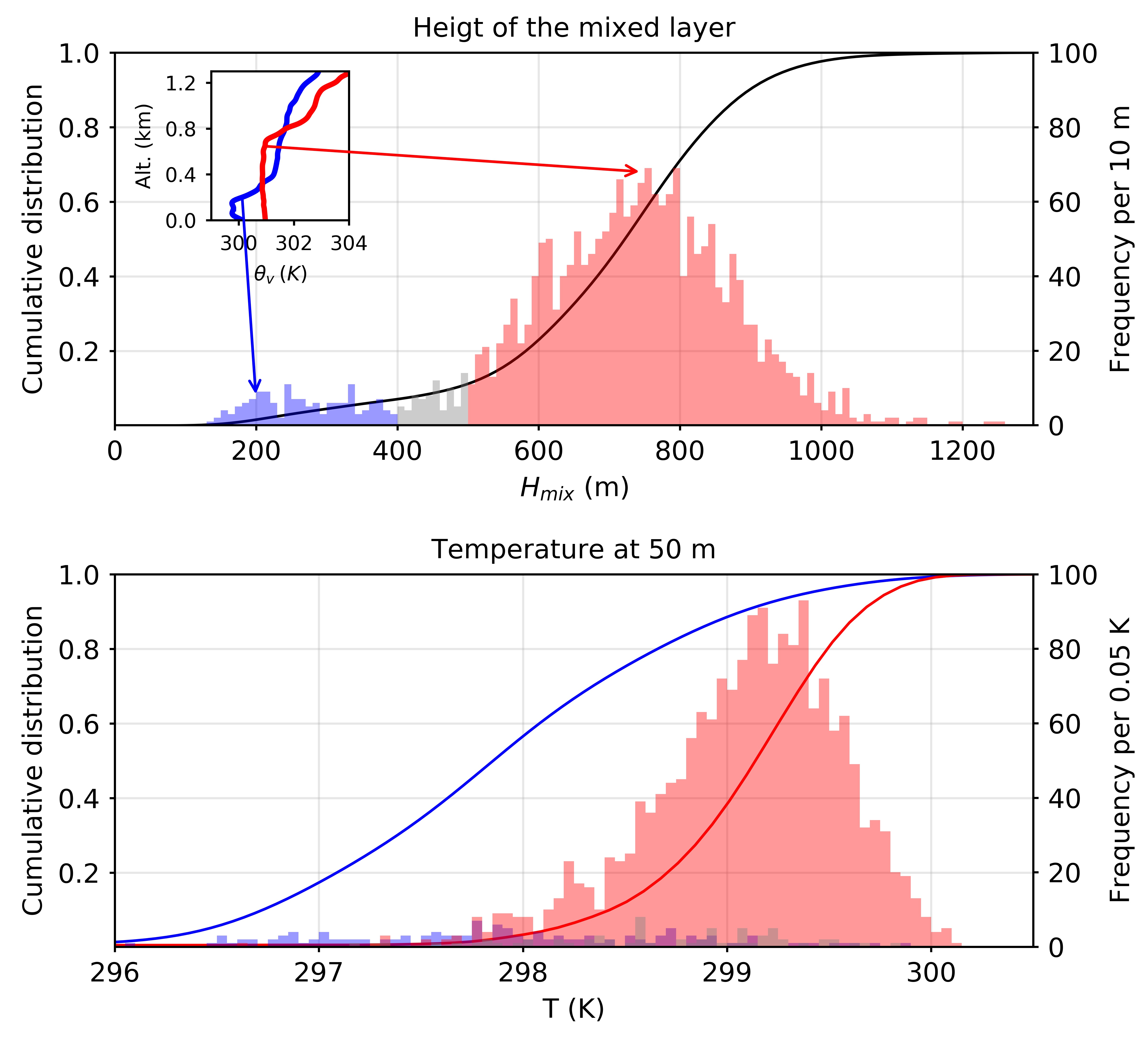}
        \caption{(top) Cumulative distribution function (black) and histogram for the height of the mixed layer for all radiosondes and dropsondes launched during EUREC$^{4}$A. The bins of the histogram are 10 m wide. Colors indicate cold pool soundings (blue), unclassified soundings (grey) and environmental soundings (red). The upper left panel shows two examples of $\theta_v$ profiles, one in a cold pool and the other in the environment. (bottom) Cumulative distribution functions for cold pool (blue) and environmental (red) temperatures at 50 meters, and the corresponding histograms with a bin width of 0.05 K.} 
        \label{dist_hmix}
\end{figure}

Figure \ref{dist_hmix} shows the $H_{mix}$ distribution for EUREC$^4$A soundings. The histogram reveals a negatively skewed distribution, with a median of 720 m. Assuming that the left tail of the distribution is due to cold pools, we choose to define 'cold pool soundings' as those with $H_{mix}$ lower than 400 m (7\% of soundings, in blue), and  'environmental soundings' as those with $H_{mix}$ higher than 500 meters (90\% of soundings, in red). With this definition, only 3\% of soundings (in grey) are neither in cold pools nor in the environment. The blue part of the distribution peaks slightly above 200 m, consistent with the typical depth of cold pools observed by \citet{Zuidema2012} in trade-wind regimes. Since this value is more than three times smaller than the typical mixed layer height in the region (Fig. \ref{dist_hmix}), the mixed layer height seems \textit{a priori} a suitable criterion to distinguish cold pools from their environment during EUREC$^4$A. In the cold pool recovery process, the height of the mixed layer gradually increases due to surface sensible and latent heat fluxes and entrainment warming and drying from above, as documented by \citet{Zuidema2012} during RICO. It could explain the large spread of the $H_{mix}$ distribution below 400 m (in blue) and its smooth increase from 400 to 500 m (in grey).

The surface temperature distribution (bottom panel, Fig. \ref{dist_hmix}) reveals that cold pool soundings are on average colder than environmental soundings. Nevertheless, a fixed surface temperature threshold -- as done for instance in \citet{Szoeke2017} with radiosondes launched during DYNAMO -- would not be able to isolate cold pools. Indeed, given the spatial and temporal extent of EUREC$^4$A, cold pools are not the only source of variability for surface temperature. In particular, measurements from research vessels reveal a sea surface temperature (SST) difference of more than 1 K between the northeast (50$^\circ$W, 16$^\circ$W) and the southwest (12$^\circ$N, 59.5$^\circ$W) of the Tradewind Alley during EUREC$^4$A. Mesoscale variability in SST features was also observed in the Boulevard des Tourbillons during the field study \citep{Stevens2020overview}. Due to this variability, the surface temperature distributions for cold pool and environmental soundings overlap (as shown in Fig. \ref{dist_hmix}) and do not allow to clearly separate cold pools from their environment. 

To show that $H_{mix}$ provides on the contrary a robust detection of cold pools, we first test our detection method in a high-resolution simulation over the Atlantic Ocean, upstream of Barbados. This simulation was conducted using the large-eddy simulation version of the ICOsahedral Non-hydrostatic (ICON) model, with realistic boundary conditions, 313-m horizontal grid spacing and 150 vertical levels  \citep{Zangl2015, Dipankar2015}. It was initialized on 11 December 2013 at 0900 UTC with 1.25km-resolution runs \citep[see][for further details]{Vial2019}. 

In high-resolution models, most cold pool detection methods rely on anomalies in surface temperature (or related quantities such as potential temperature or buoyancy), winds, or surface rain \citep[see][and references therein]{Drager2017}. To determine cold pool edges, they usually requires the entire 2 or 3D field of some thermodynamic or dynamic variables and are therefore unusable with sounding data. On the contrary, our detection method is based on 1D thermodynamic profiles and is thus entirely local. 



\begin{figure}[t]
        \noindent\includegraphics[width=\textwidth]{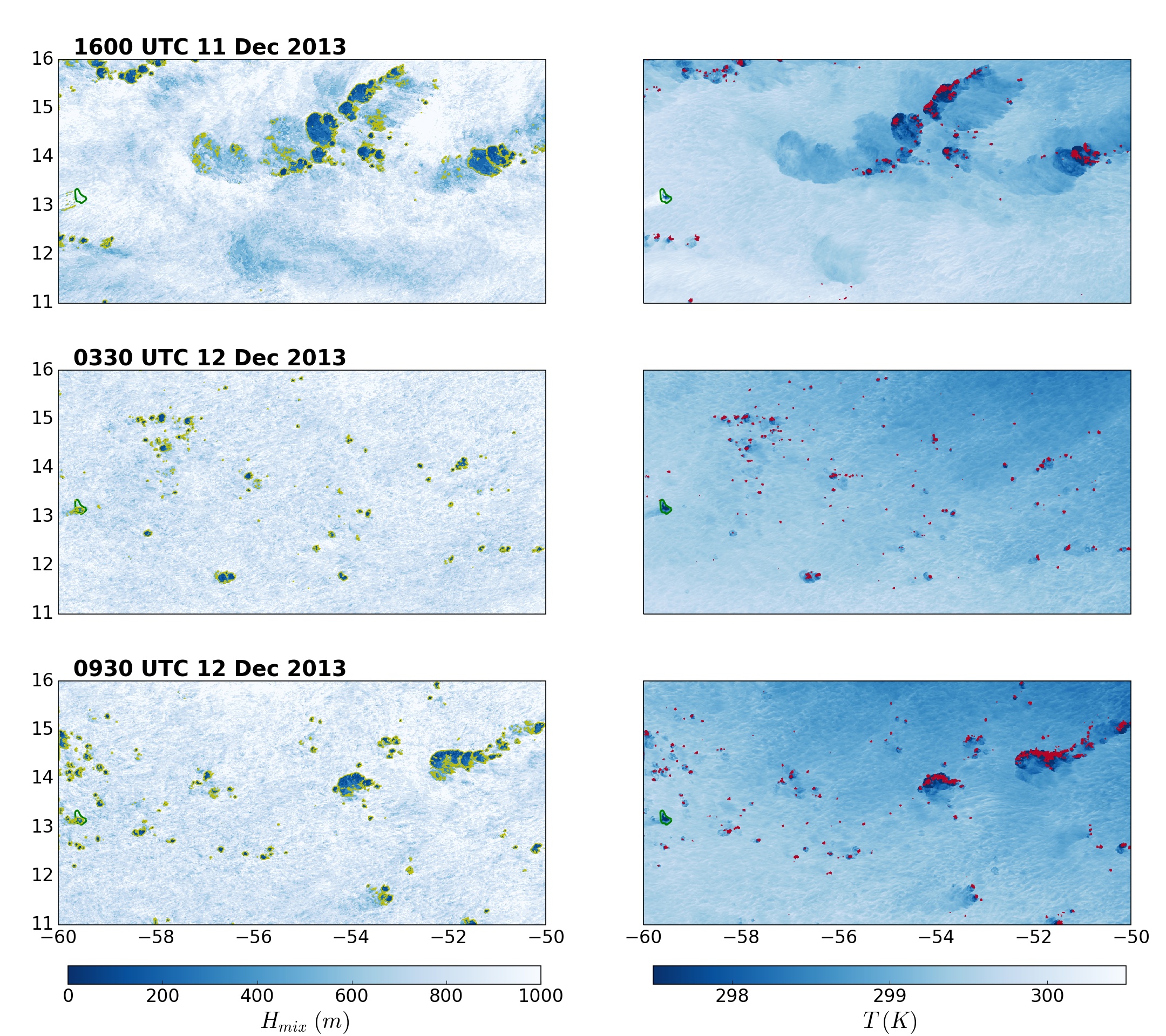}
        \caption{Height of the mixed layer $H_{mix}$ and surface temperature T at $z \approx 50 \: m$ in an area of $5^\circ {\times} 10^\circ$ upstream of Barbados (circled in green) on December 11, 2013 at 1600 UTC and December 12, 2013 at 0330 UTC and 0930 UTC. In the left-hand panels, the red dots represent locations where surface rainfall is greater than \text{10 mm day$^{-1}$}. In the right-hand panels, cold pools are circled in yellow as regions where $H_{mix}$ is lower than 400 meters. 
    } 
        \label{ICON}
\end{figure}

Despite this simplicity, Figure \ref{ICON} shows that it gives consistent results with the surface temperature and precipitation fields. It represents three snapshots of the mixed layer height $H_{mix}$ (left) and the temperature near the surface ($z \approx 50 \: m$) (right) on 11 December 2013 at 1600 UTC and on 12 December 2013 at 0330 UTC and 0930 UTC. In the right panels, we see that regions with negative temperature anomalies are co-located with significant surface precipitation (red dots) and also with important wind shifts (not shown), suggesting that these cold regions are in fact convective cold pools. In the left panels, we apply our detection method by circling in yellow regions where the mixed layer is less than 400 meters. Qualitatively, there is a really good agreement between cold pools detected with our method and regions strongly cooled by rainfall. $H_{mix}$ is unaffected by large-scale temperature gradients and seems to discriminate cold pools everywhere in the domain. Conversely, precipitation alone is a not a distinguishing factor of cold pools, as it is generally present only over a small part of cold pools, and any method based on a temperature threshold would likely be strongly sensitive to the temperature gradient between the northeast and the southwest of the domain. It gives us confidence to choose $H_{mix}$ in the next section to detect cold pools from soundings during EUREC$^4$A. 

\section{Application to EUREC$^4$A data}

\label{results}

\begin{figure}[t]
        \includegraphics[width=\textwidth]{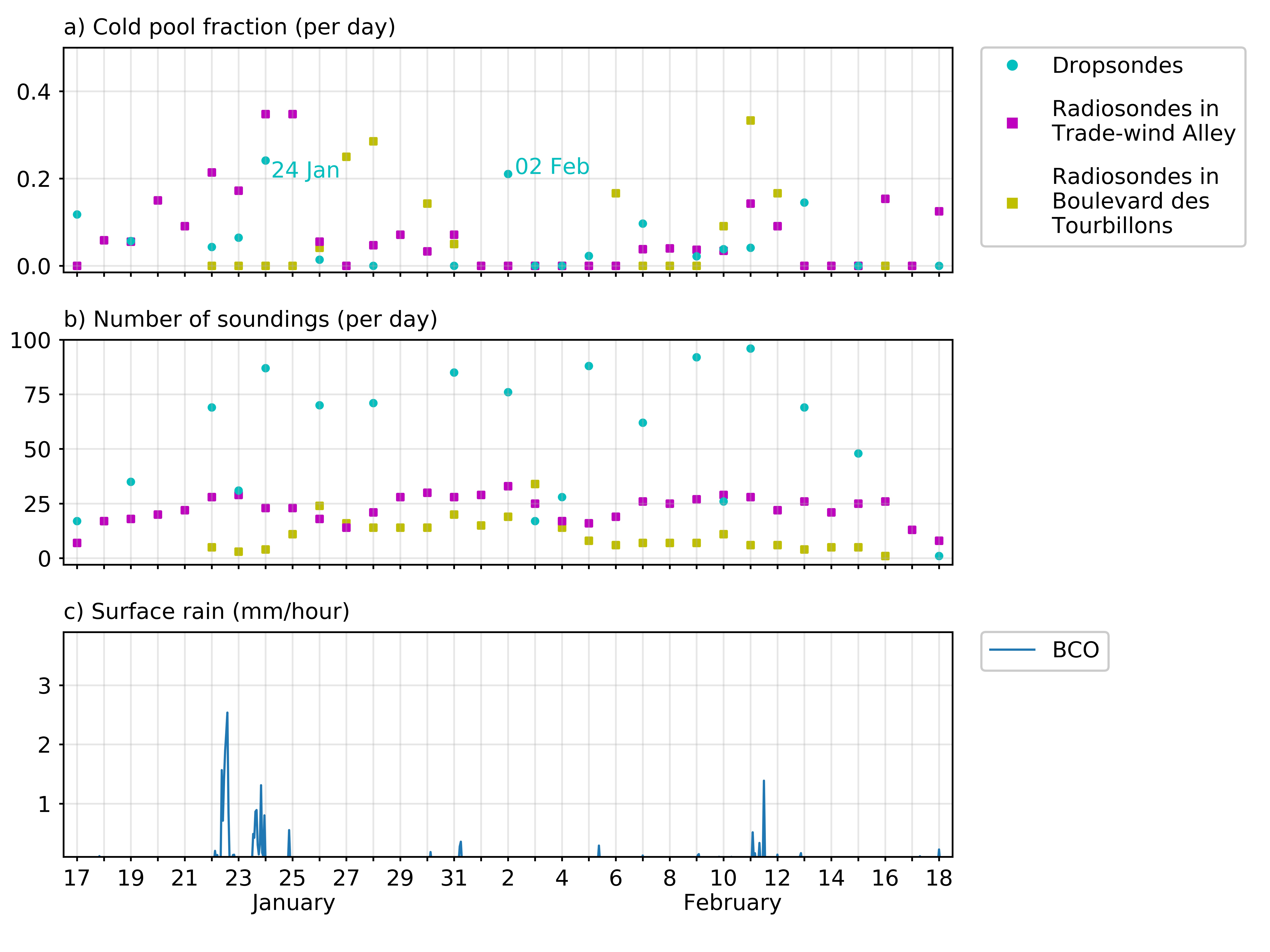}
        \caption{(a) Cold pool fraction per day from dropsondes (cyan circles), and from radiosondes launched in the Trade-wind Alley (magenta squares) and in the Boulevard des Tourbillons (yellow squares). The two days selected for the case studies are indicated in cyan. (b) Number of soundings per day, with the same color code. (c) Hourly rain rate at BCO from the micro-rain radar at 325\,m height \citet[see][for details]{Stevens2016}.
}
        \label{evolution_cp_campaign}
\end{figure}

Applying our method to our set of 2174  EUREC$^4$A soundings, we find that 149 sondes have fallen into cold pools. Figure \ref{evolution_cp_campaign} shows (a) the cold pool fraction, (b) the number of soundings per day and (c) the hourly surface rain rate measured at BCO. As expected, we see some consistency between rain rates measured at BCO and cold pool fraction in dropsondes and radiosondes in the Trade-wind Alley, in particular around January 22-25 and February 11-12, when significant rainfall rates were measured at BCO. For the dropsondes, the largest cold pool fractions have been observed on January 24 and February 2. In the following, we will examine the cold pool characteristics on these two days in more detail.

 \subsection*{January 24}

\begin{figure}[t]
        \centering
        \includegraphics[width=0.7\textwidth]{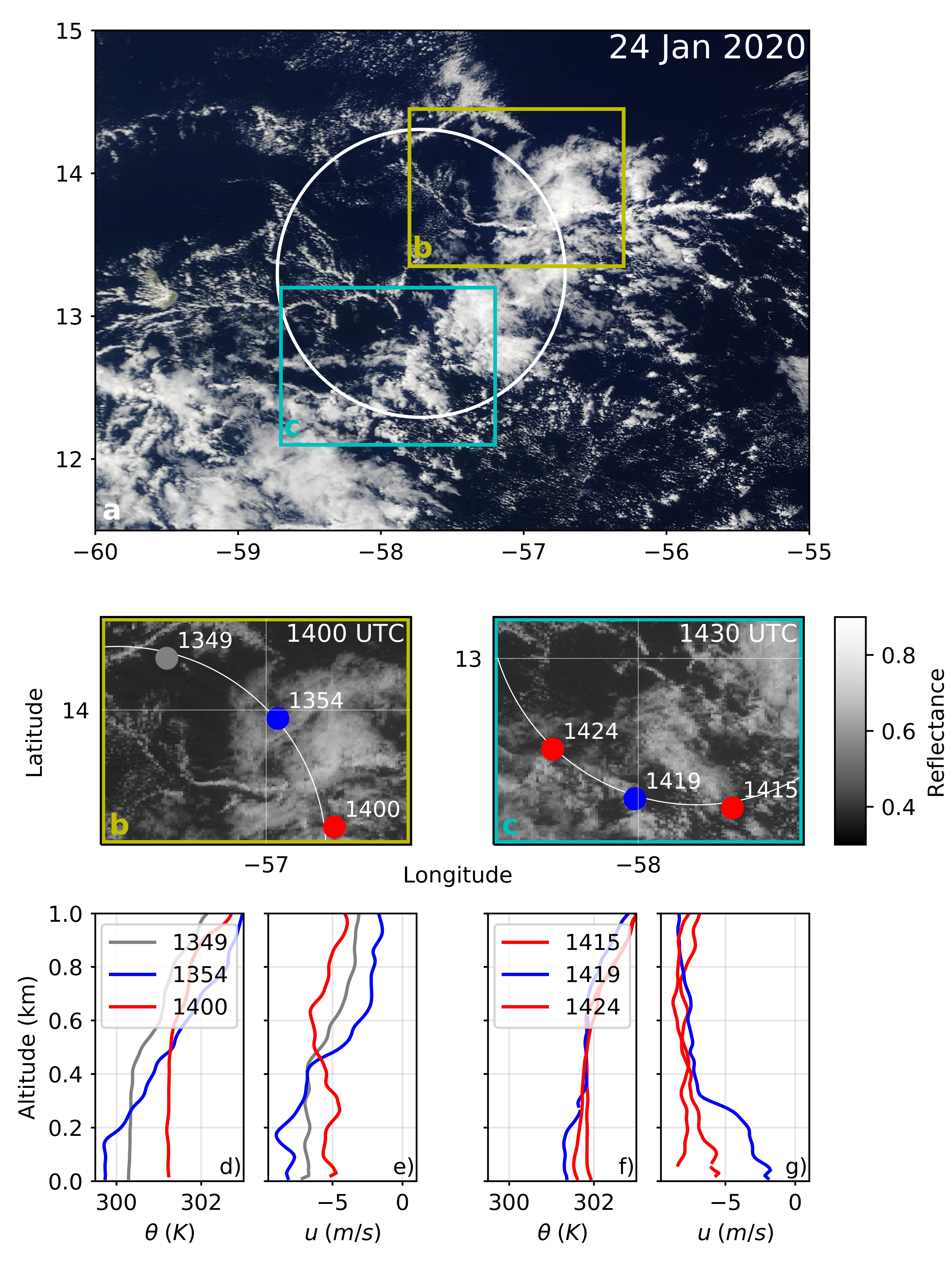}
        \caption{(top) MODIS-Terra scene from Worldview upstream of Barbados on January 24, 2020. The EUREC$^4$A circle is indicated in white. The rectangles mark the two regions shown in the lower panels. (middle) GOES-16 visible reflectance (channel 2) displayed at 1400 UTC (b) in the northeast quarter of the EUREC$^4$A circle and at 14:30 UTC (c) in the southwest part of the circle. The ground position and launch time of dropsondes dropped in the 15 minutes preceding the satellite image is shown in blue for cold pool soundings, in red for environmental soundings and in grey for unclassified soundings. (bottom) Profiles of potential temperature (d-f) and zonal wind speed (e-g) from the surface to 1 km for the dropsondes highlighted in the middle panel.}
        \label{20200124}
\end{figure}

On January 24, more than 20 \% of dropsondes (18 out of 88, see Fig. \ref{evolution_cp_campaign}) launched by HALO fell into cold pools. The MODIS-Terra satellite image of that day (Fig. \ref{20200124}) reveals the presence of many mesoscale arcs. As noted by previous observational and modelling studies of trade-wind regimes \citep{Zuidema2012, Seifert2013}, such mesoscale arcs are likely due to cold pools. Indeed, while the dense air inside cold pools is less able to support buoyancy-driven convection, it has long been known that the edges of an expanding cold pool can lift surrounding air masses and trigger new convective cells \citep[e.g.][]{Purdom1976, Knupp1982, Weisman2004}. Based on these previous studies, we expect relatively few clouds above cold pools, but mesoscale arcs of clouds near their border. The presence of many mesoscale arcs around the EUREC$^4$A circle is thus consistent with the numerous cold pools detected that day from the HALO dropsondes. 

Superposing the launch positions of dropsondes to GOES-16 visible channels reveals the cloud field dropsondes have actually sampled. Panel b (yellow box) shows the position of dropsondes launched by HALO when passing over a mesoscale arc in the northeastern part of EUREC$^4$A circle, just before 1400 UTC. The bottom panels (d-e) give the potential temperature and zonal wind speed measured by these dropsondes. The cold pool sounding (blue) is the coldest, with the strongest zonal wind near the surface and the shallowest mixed layer. In terms of potential temperature, $H_{mix}$ and zonal wind, the unclassified sounding (grey) represents an intermediate case between the cold pool and the environmental (red) soundings. Located inside the mesoscale arc, but closer to its edges than the cold pool sounding, it suggests that the mixed layer may partially recover its initial properties near the edge of a cold pool. 

Panel c (cyan box) gives a second example of a cold pool sounding. In this example, due to the presence of many clouds on the satellite image, it is hard to say whether the cold pool sounding fell in a cloud arc or not. Our detection method seems nevertheless consistent with wind measurements, which reveal the presence of a cold outflow at the surface (with a difference of more than 3 m s$^{-1}$ in near-surface zonal velocity between the blue and red soundings). Note that in this example, the surface temperature does not distinguish cold pools, as the blue sounding (1419 UTC) in panel c is less than 0.5 K colder than the neighboring red soundings (1415 and 1424 UTC), and slightly warmer than the red sounding in panel b (1400 UTC).

\subsection*{February 2}

\begin{figure}[t]
        \centering
        \includegraphics[width=0.8\textwidth]{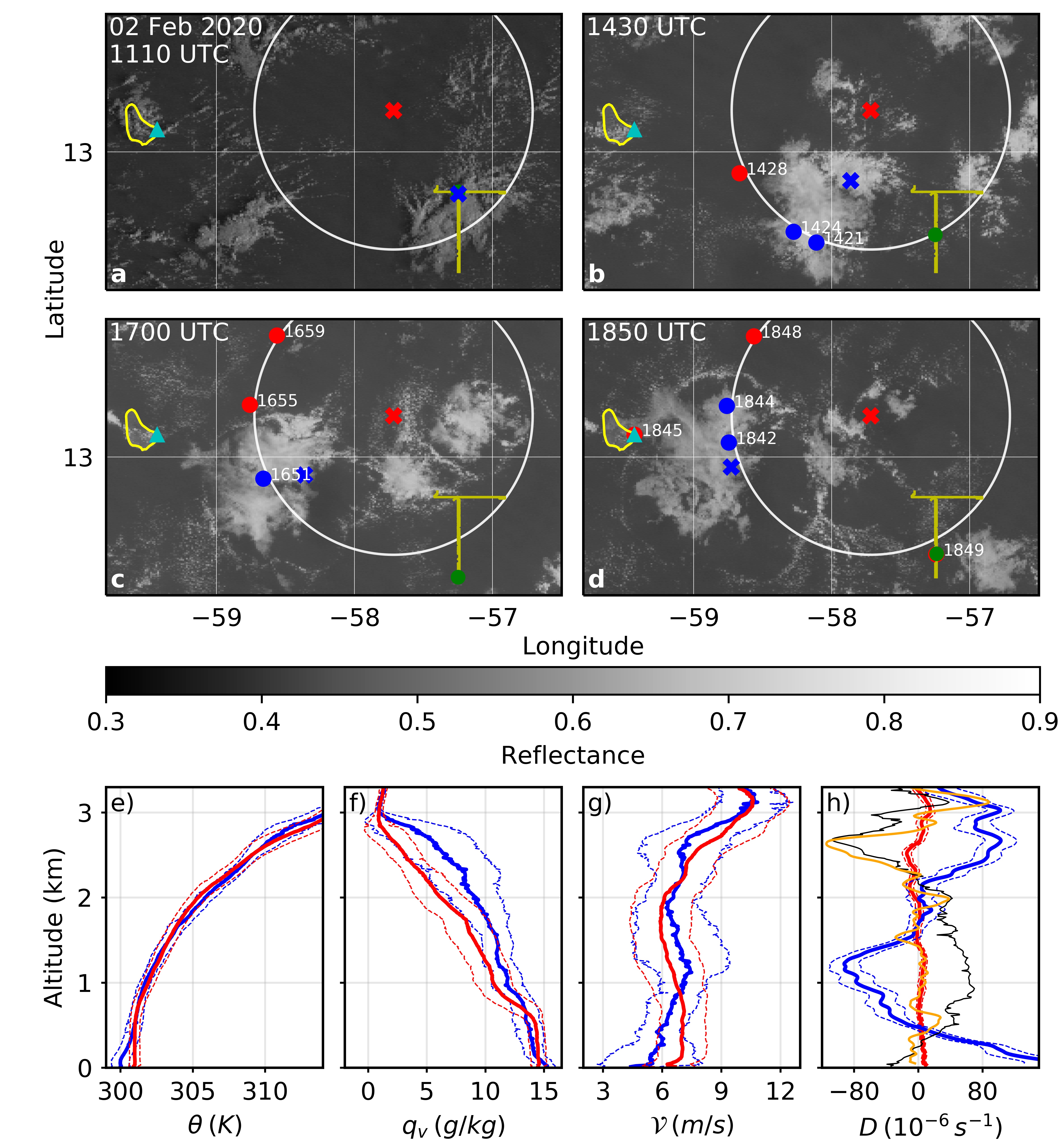}
        \caption{(top) GOES-16 visible reflectance (channel 2), displayed on February 2 at (a) 1130, (b) 1430, (c) 1700 and (d) 1850 UTC in the Atlantic region upstream Barbados (circled in yellow). The EUREC$^4$A circle is outlined in white. The ground position and launch time of dropsondes dropped in the 15 minutes preceding the satellite image is shown in blue for cold pool soundings and in red for environmental ones. The yellow line represents the path of the Meteor, with green-filled circle marking the position of the vessel at the time the satellite image was taken. The location of BCO is indicated by a cyan triangle. (bottom) (e) Mean potential temperature, (f) specific humidity, (g) wind speed and (h) divergence in cold pool and environmental soundings. In each panel, the standard deviation around the mean for the two types of soundings is represented by a dashed line. In panel h, the black line indicates an example of divergence calculated over an entire EUREC$^4$A circle (from 1841 to 1936 UTC) using all dropsondes (12 in total, including 4 in cold pools), and the orange line indicates the divergence over the same circle by considering environmental dropsondes only. For the sake of clarity, the x-axis is stretched by a factor of 4 for these two curves, that is one should read $ \pm 20$ instead of $\pm 80$ ($10^{-6}s^{-1}$).}
        \label{20200202}
\end{figure}

\begin{figure}[t]
        \includegraphics[width=\textwidth]{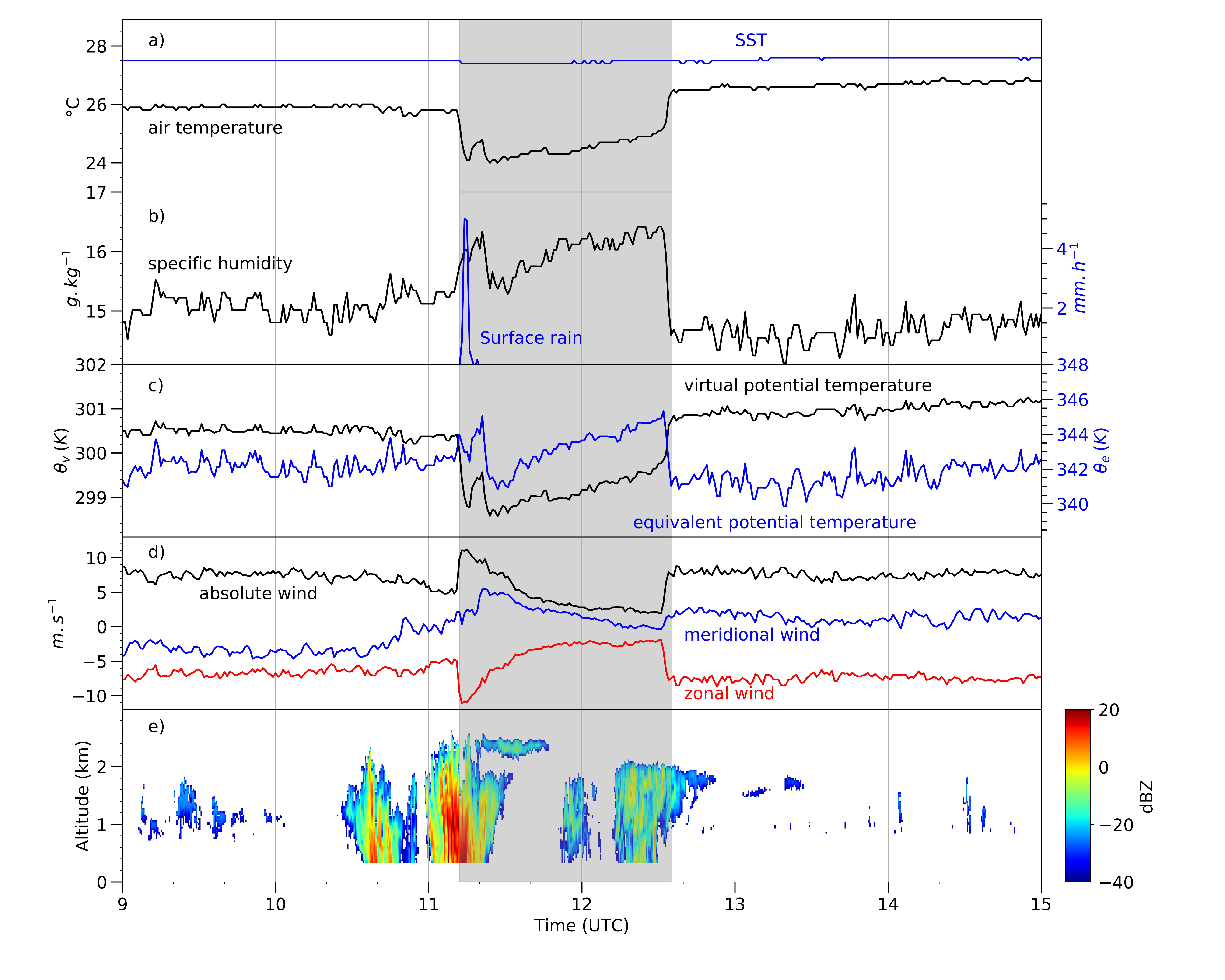}
        \caption{Shipboard measurements from the Meteor on February 2, from 0900 to 1500 UTC: (a) surface air (black) and sea temperature (blue), (b) surface specific humidity (black) and surface rain rate (blue), measured using a ship rain gauge SRM 450. (c) virtual potential temperature (black) and equivalent potential temperature (blue), (d) Zonal (red), meridional (blue) and absolute (black) wind speed, (e) Vertically pointing shipboard W-band radar data at 94 GHz, measured using the Raman lidar system LICHT (Lidar for Cloud, Humidity and Temperature profiling) formerly at BCO and described in \citet{Stevens2016}. Color bar indicates uncalibrated radar signal-to-noise ratio, best interpreted relative to itself. In the five panels, the grey shading marks the cold pool time period. Meteor data are freely accessible at https://observations.ipsl.fr/aeris/eurec4a/.}
        \label{Meteor}
\end{figure}

On February 2, the satellite images reveal a cloud field with many "flowers", characterized by the presence of mesoscale, quasi-circular, stratiform shallow clouds, capped by a strong inversion and separated by very dry clear-sky areas following the classification introduced in \citet{Stevens2019}. At 1700 and 1850 UTC, in the eastern part of EUREC$^4$A circle, a mesoscale cloud arc is visible in all directions around one of the flowers. As on January 24, we suspect this cloud arc to be due to the presence of a cold pool at the surface. This claim is supported by the surface meteorological data from the research vessel Meteor, which captured the onset of the corresponding cold pool around 1110 UTC (see Fig \ref{20200202}, panel a). Indeed, as shown in Fig. \ref{Meteor}, the surface air temperature (a) measured by the Meteor at 1110 UTC suddenly dropped by approximately 1.5$^\circ$C, suggesting that the research vessel entered a cold pool. This drop is also visible in the virtual potential temperature (c). The ship remained in the cold pool until 1230 UTC when the temperature rose to 26.2$^\circ$C, a slightly higher value than before the passage of the cold pool. The wind speed (d) changed significantly as the ship passed through the cold pool front: the absolute wind speed increased rapidly from $5 \textrm{ m s}^{-1}$ to $10 \textrm{ m s}^{-1}$ at the time of the temperature drop, before decreasing continuously to $2 \textrm{ m s}^{-1}$ and going back to its pre-cold pool value of $5 \textrm{ m s}^{-1}$ after the cold pool passed. The wind speed variations upon entering and leaving the cold pool thus have approximately the same amplitude. These results stand both for the meridional and the zonal wind, although the wind shifts have slightly stronger amplitudes for the zonal wind. The specific humidity (b) and the equivalent potential temperature (d) reveal that this cold pool is overall moister than its environment, especially near its edges, which are about 1 g kg$^{-1}$ moister than its core. Finally, the W-band radar (e) shows 2.5 km deep clouds with strong reflectivity near the edges of the cold pool. When the ship enters the cold pool, these clouds coincide with a high surface rain rate of more than $4 \textrm{ mm h}^{-1}$ (b), so we suspect the evaporation of precipitation below these clouds to feed this cold pool. Overall, these different features are consistent with previous observations of cold pools over tropical oceans \citep{Zuidema2012, Szoeke2017} and support the cold pool tagging shown in Fig. \ref{20200202}.

Over the 100 dropsondes and radiosondes launched near the EUREC$^4$A circle that day, 16 were detected falling in cold pools. 13 of them fell in or around the cloud cluster passing over the Meteor at 1110 UTC. The three other cold pool soundings all fell in or around other cloud clusters, located further east of the EUREC$^4$A circle (visible for example at 1700 UTC in Fig. \ref{20200202}). Mean sounding profiles in (blue) and out (red) of cold pools are shown in the bottom panels (e-g). Cold pool soundings are on average 1 K colder than the other soundings in terms of potential temperature. This difference is mainly observed below 400 meters. The air in the cloud layer above cold pools, from 800 m to 3 km, is significantly moister than the ambient air, consistent with the presence of cloud clusters on satellite images. In the SCL, we observe on the contrary slightly drier air between 200 meters and 800 meters in cold pools compared to their environment, which might be due to downdrafts transporting the relatively drier air from the cloud layer to the sub-cloud layer above cold pools. On this day, near-surface air in cold pools is on average moister than in their environment, consistent with Meteor observations. At the surface, cold pools propagate as density currents and induce significant wind shifts, visible in the relatively large standard deviation of the near-surface wind in cold pools (g). The consistency between satellite images, Meteor data and the thermodynamic profiles derived from soundings support the ability of our detection method to distinguish cold pools from their environment on February 2.

This robust distinction between cold pool and environmental soundings allows to study not only the properties of cold pools, but also those of the convective system to which they belong. For instance, the longevity of the cloud cluster first detected at 1110 UTC by the Meteor (Fig \ref{Meteor}) and still visible on satellite images at 1850 UTC (Fig \ref{20200202}) suggests intense convective activity above the cold pool detected in the southern part of the EUREC$^4$A circle. To characterize the associated air circulations, we compute the divergence in and above this cold pool, and compare it to the divergence calculated over the EUREC$^4$A circle from environmental soundings. For that, we use the method introduced by \citet{Lenschow2007} and successfully tested during the Next Generation Aircraft Remote Sensing for Validation Studies \citep[NARVAL2,][]{Bony2019}. This method assumes that the wind field is stationary and that wind variations in longitude and latitude are linear at each vertical level, defining thus the large-scale wind $\boldsymbol{V}=(u, v)$ such that:\begin{equation}
    \boldsymbol{V} = \boldsymbol{V}_0 + \frac{\partial \boldsymbol{V}}{\partial x} \Delta x + \frac{\partial \boldsymbol{V}}{\partial y}\Delta y
    \label{wind}
\end{equation}
where $\boldsymbol{V}_0$ is the mean velocity over the area and $\Delta x$ and $\Delta y$ are the eastward and northward displacements from a chosen center point. Since $\boldsymbol{V}$, $\boldsymbol{V}_0$, $\Delta x$ and $\Delta y$ are known, $\frac{\partial \boldsymbol{V}}{\partial x}$ and $\frac{\partial \boldsymbol{V}}{\partial y}$ can be calculated using a simple least square fit in equation \ref{wind} and the divergence can then be derived as: $D=\partial_x u + \partial_y v$.

As in \citet{Bony2019}, for HALO dropsondes out of cold pools, we choose as center point the center of the EUREC$^4$A circle, indicated by a red cross in Fig. \ref{20200202}. For cold pool soundings, the center point is taken at the center of the cold pool visible in the southern part of the EUREC$^4$A circle. To estimate the center location, we take as a starting point the position of the Meteor where the minimum temperature was measured and advect it by the surface mean wind speed measured by the research vessel in a four-hour time slot centered on the cold pool period. In Fig. \ref{20200202}, the position of the estimated "cold pool center" is indicated in each snapshot by a blue cross. We see that it matches approximately the center of mesoscale arcs observed at 1700 and 1850 UTC. With respect to, respectively, the center of the EUREC$^4$A circle and the estimated center of the cold pool, we then compute the least square fit for the 29 dropsondes dropped in the EUREC$^4$A circle out of cold pools from 1420 to 1845 UTC, and for the 11 dropsondes dropped in the main cold pool during the same period. This period is long enough to have a sufficient number of dropsondes to get a robust measure of divergence in and out of cold pools, but short enough to consider the stationarity assumption as valid at least out of cold pools, as the autocorrelation time-scale of large-scale divergence estimated by \citet{Bony2019} is approximately 4 hours. Assuming stationarity of the wind field above the cold pool is more controversial, but seems nevertheless a good approximation in this precise situation due to the duration of the convective system and the coherence between dropsonde measurements at different times.

The blue curve in Fig. \ref{20200202} h reveals a layer of strong divergence near the surface in the cold pool, culminating at more than $80 \times 10^{-6} \textrm{ s}^{-1}$ close to the ground, consistent with the spreading of the cold pool at the surface. On the opposite, from 500 m to 1.5 km, we observe a zone of convergence peaking slightly above 1 km with a value close to $-80 \times 10^{-6} \: \textrm{ s}^{-1}$. This converging air might feed convective updrafts or downdrafts near cloud base. Finally, between 2 and 3 km, there is again a divergent layer, with a magnitude of about $80 \times 10^{-6} \textrm{ s}^{-1}$. It corresponds to the altitude where the updraft air is detrained and where the stratiform outflow layer is located, though the divergence profile close to the inversion might also be influenced by local circulations induced by mixing with drier free tropospheric air. In contrast, the divergence plot out of cold pools has a much smaller amplitude from the surface to the inversion. As expected, above the inversion, the divergence patterns in and out of cold pools follow each other closely (not shown), suggesting homogeneous large-scale wind patterns in the free troposphere, whether or not there are cold pools below.

The mesoscale divergence values above the cold pool are two or three times larger than the maximum large-scale divergence observed by \citet{Bony2019} during NARVAL2 over 180 km diameter circles. When averaged over a EUREC$^4$A circle, they could therefore leave an imprint on the large-scale divergence, as shown by the black line which represents the divergence computed over a EUREC$^4$A circle from 1841 to 1936 UTC using all dropsondes (12 in total, including 4 dropped in cold pools). The orange line shows the divergence calculated over the same circle with environmental soundings only. The comparison between the orange and black curves shows local differences of more than $10 \times 10^{-6} \textrm{ s}^{-1}$ between the two curves. The largest differences are observed between 500 m and 2 km, with smaller differences also present near the surface and just below the inversion, at 3 km. The differences between the black and orange curves are consistent with the differences between the blue and red ones, albeit of opposite sign: indeed, as the cold pool center is located outside the EUREC$^4$A circle at 1850 UTC (see the blue cross on Fig. \ref{20200202}, panel d), convergence over the cold pool at that time leads to divergence over the EUREC$^4$A circle and vice versa. This example shows that mesoscale circulations around cloud clusters such as the one observed here could explain part of the variability of the large-scale divergence in the SCL and cloud layer noted in \citet{Bony2017}.

In large-eddy simulations of marine shallow cumulus convection, \citet{Bretherton2017} found circulations similar to those observed around the main flower on February 2. According to the authors, these circulations might participate in the mesoscale aggregation of shallow convection below the trade-inversion. Indeed, in trade-wind regimes, \citet{Bretherton2017} show that such circulations induce relative moistening of the moistest sub-domains, a form of gross moist instability. This hypothesis, if verified, would give a possible explanation for the persistence for several hours of the cloud clusters on February 2 despite the presence of cold pools below them. 

\subsection*{Statistics for other days}

\begin{figure}[t]
        \includegraphics[width=\textwidth]{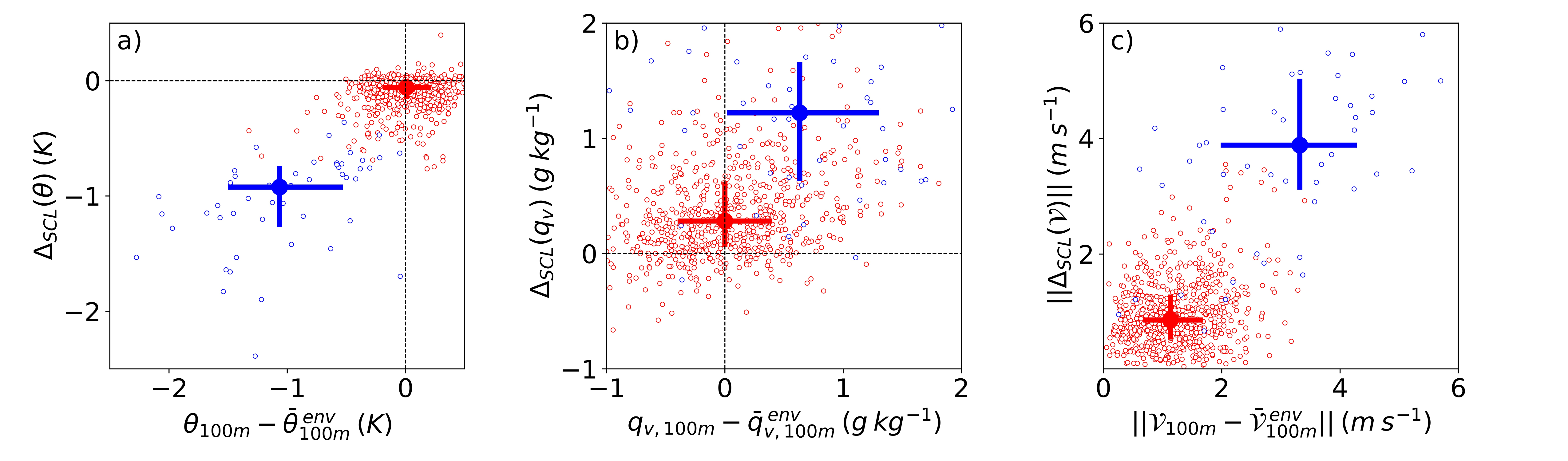}
        \caption{Scatter plot of the vertical variations of (a) potential temperature, (b) specific humidity and (c) wind speed between 100 and 500 m vs spatial anomalies of the same variables at 100 m. The soundings considered are dropsondes launched in EUREC$^4$A circles, and spatial anomalies are calculated with respect to the circle environmental mean, that is the mean calculated circle by circle with dropsondes launched out of cold pools only. Also reported are the median values for cold pools (blue) and their environment (red). The bars indicate the 25th and 75th percentiles of the distributions.}
        \label{test_HALO_sondes}
\end{figure}

We will now look at a few statistics of cold pools for other days during EUREC$^4$A, focusing first on the HALO dropsondes launched in the EUREC$^4$A circles. In total, the aircraft dropped more than 700 dropsondes along 72 circles, that is about 10 dropsondes per circle. We use circle sounding data to compare the characteristics of cold pools to those of their close environment. In Fig. \ref{test_HALO_sondes}, the x-axis represents the difference between the potential temperature (a), specific humidity (b) and wind speed (c) measured by each dropsonde at $100 \textrm{ m}$ and the mean value at the same altitude for environmental dropsondes launched in the same EUREC$^4$A circle. Mathematically, we define in the following $X_{z_0}=\frac{1}{100m}\int_{z_0-50m}^{z_0+50m}X(z)dz$ as the value of $X$ at the altitude $z_0$ averaged over a 100 m depth and $\bar{X}_{z_0}^{env}$ as the mean circle value of $X_{z_0}$ in environmental soundings. We average $X_{z_0}$ over a 100 m interval in order to be less sensitive to local vertical heterogeneities. With these notations, the x-axes become respectively $\theta_{100m} - \bar{\theta}_{100m}^{env}$, \text{$q_{v,100m} - \bar{q}_{v,100m}^{env}$} and \text{$\lVert \bm{\mathcal{V}}_{100m}-\bm{\mathcal{V}}_{100m}^{env} \rVert$}. They show that cold pools are on average 1 K colder and \text{1 g kg$^{-1}$} moister than their environment, and that the wind difference between individual soundings and mean circle values is on average 2 m s$^{-1}$ larger for cold pools than for their environment.

The y-axis quantifies the imprint of cold pools on the vertical profiles for the same variables. For each sounding, the y-coordinate represents the difference $\Delta_{SCL}(X)$ in the SCL between 100 and 500 m, that is in and above cold pools when there is one. Mathematically, we define $\Delta_{SCL}(X) = X_{100m} - X_{500m}$ for both cold pool and environmental soundings. In the environment, on average, there is little difference in terms of potential temperature, specific humidity and wind speed between the two layers, consistent with what we would expect for a relatively well-mixed layer. On the opposite, cold pool soundings are on average 1 K cooler and 1 g kg$^{-1}$ moister at 100 m, and the average wind difference is about 3.5 m s$^{-1}$ between the two layers. These values are consistent with those of the x-axis and suggest that in a well-mixed layer the vertical imprint of cold pools can be used to estimate the differences between cold pools and their environment in terms of potential temperature, specific humidity and wind speed.  

\begin{figure}[t]
        \includegraphics[width=\textwidth]{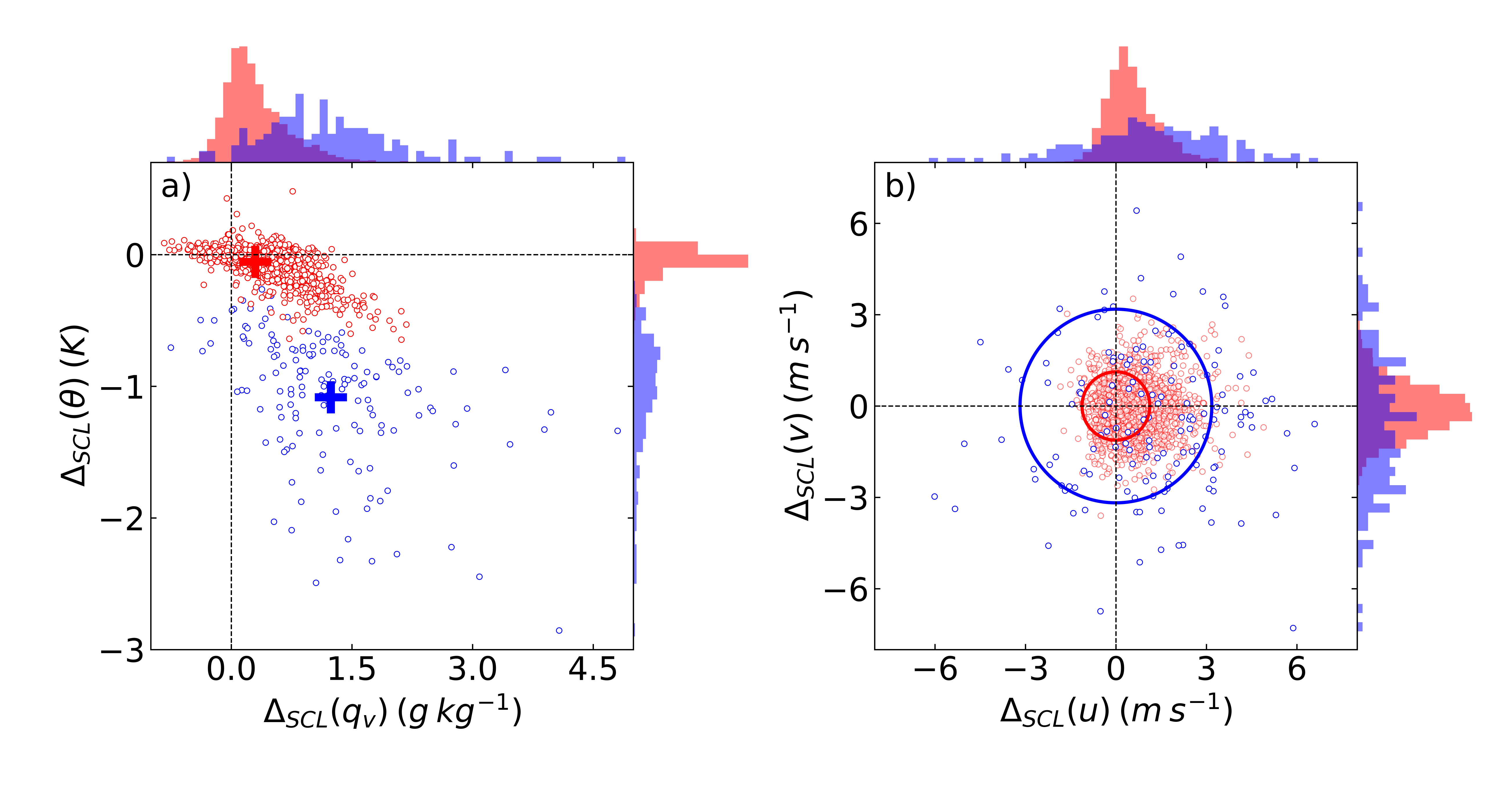}
        \caption{(left) Scatter plot of the difference of potential temperature at 100 m and at 500 m $\Delta_{SCL}(\theta)$  vs difference of specific humidity between the same altitudes $\Delta_{SCL}(q_v)$ for cold pool (blue) and environmental (red) soundings.  The crosses indicate the mean of each distribution. On the upper and right sides of the graph, marginal distributions of each variable are also reported for cold pools and their environment. The bins of the histograms are $0.1 \textrm{ g kg}^{-1}$ wide for the specific humidity and $0.1 \textrm{ K}$ wide for potential temperature. To facilitate the comparison, the area under each distribution has been normalized to 1. (right) Same plot, but for the meridional and zonal winds, with a bin width of 0.3 m s$^{-1}$ for the corresponding histograms. The circles represent the mean of the absolute wind speed distributions, that is the mean of $\sqrt{(\Delta_{SCL}(u))^2 + (\Delta_{SCL}(v))^2}$ for cold pool and environmental soundings.
        }
        \label{var_vs_var}
\end{figure}

\begin{figure}[t]
        \includegraphics[width=\textwidth]{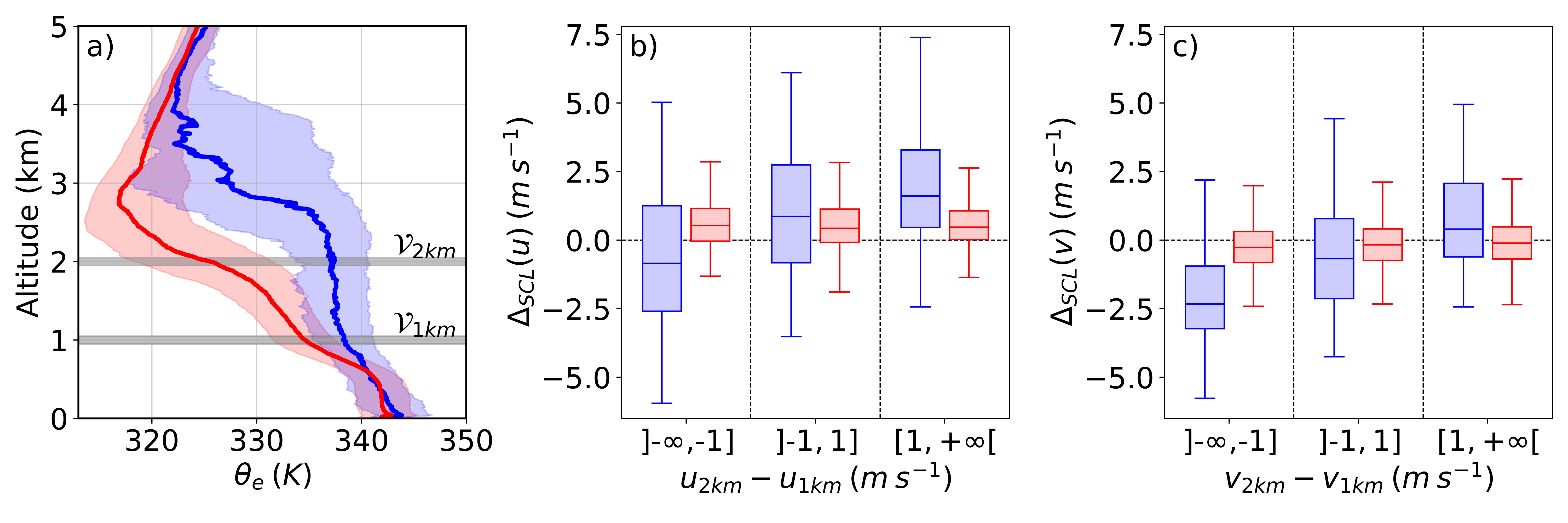}
        \caption{(a) Profiles of equivalent potential temperature for all EUREC$^4$A soundings launched in cold pools (blue) and in their environment (red).  The bold lines represent the medians of each distribution and the color shading their interquartile range. The grey shading indicate the altitudes taken to define $\mathcal{\textbf{V}}_{1km} = (u_{1km},v_{1km})$ and  $\mathcal{\textbf{V}}_{2km} = (u_{2km},v_{2km})$ used in the middle and right panels. (b) Box and whisker plots for the zonal wind difference between 100 and 500 m  $\Delta_{SCL}(u)$ for three zonal wind shear ranges in the cloud layer. The wind shear ranges are defined by calculating for each sounding the difference between $u_{2km} = \int_{1.8km}^{2.2km} u(z) dz$ and $u_{1km} = \int_{0.8km}^{1.2km} u(z) dz$. Horizontal line within each box represent the median, box bottom and top are 1st (Q1) and last (Q3) quartile of the distributions, and the whiskers extend up to 1.5 interquartile range above Q3 and below Q1. (c) Same plot, but for the meridional wind difference between 100 and 500 m $\Delta_{SCL}(v)$ for three meridional wind shear ranges in the cloud layer.
        }
        \label{whisker_plots}
\end{figure}

Based on these results, we take in Fig. \ref{var_vs_var} the vertical imprint of cold pools as a proxy to estimate the intensity of cold pool anomalies. It allows us to generate statistics for all EUREC$^4$A soundings, including the many radiosondes and dropsondes not launched in EUREC$^4$A circles for which we do not have any environmental reference to assess cold pool properties. Consistently with Fig. \ref{test_HALO_sondes}, cold pool soundings are on average 1 K colder at 100 m than at 500 m, 1 g kg$^{-1}$ moister and experience a wind difference between the two layers 2 m s$^{-1}$ larger than for environmental soundings. The coldest cold pools are also those for which the humidity difference is the most important, although potential temperature alone explains only a small part of the total humidity variance for cold pool soundings ($R^2 = 0.19$). The difference in humidity profiles in the SCL between cold pools and environment is also visible in Fig. \ref{whisker_plots} a, which gives the equivalent potential temperature in cold pool and environmental soundings. This panel further shows that cold pool soundings are well mixed in equivalent potential temperature between 1 and 2.5 km, suggesting efficient convective mixing above cold pools.

In Fig. \ref{var_vs_var} b, the zonal and meridional wind distributions show that cold pools spread in all directions, consistently with the conceptual picture of a density current propagating over a solid boundary. In the presence of vertical wind shear, theories and numerical simulations nevertheless predict that cold pool spreading may not be perfectly isotropic, as vertical momentum transport might favor the propagation of cold pools downshear \citep{Grant2020, Mahoney2009, Moncrieff1992}. Fig. \ref{whisker_plots} middle and right panels test this hypothesis by showing the zonal and meridional wind difference between 100 and 500 m ($\Delta_{SCL}(u)$ and $\Delta_{SCL}(v)$) for three zonal and meridional wind shear ranges in the cloud layer, between 1 and 2 km: $] {-\infty}; -1 ]$, $]-1, 1]$ and $]1, +\infty[$ (m s$^{-1}$). The box and whisker diagrams reveal that $\Delta_{SCL}(u)$ is gradually shifted to higher values as the zonal wind shear in the cloud layer increases.  On the contrary, in environmental soundings, the zonal wind shear in the cloud layer has little, if any, influence on $\Delta_{SCL}(u)$. The same result stands for the meridional wind, suggesting significant momentum transport by downdrafts in cold pools in all directions. $\Delta_{SCL}(u)$ is higher than $\Delta_{SCL}(v)$ for both cold pool and environmental soundings because of strong zonal winds in the region, around $-8.1$ m s$^{-1}$ at 100 m on average, compared to $-1.7$ m s$^{-1}$ at 100 m for mean meridional winds.  These strong winds create friction in the lowest layers and decrease $u_{100m}$ with respect to $u_{500m}$, thus increasing $\Delta_{SCL}(u)$.The relation between the direction of propagation of cold pools and the vertical shear, as well as the vertical imprint of cold pools on dynamic and thermodynamic profiles illustrate some interesting results inaccessible with detection methods based on time series \citep{Young1995, Zuidema2012, Szoeke2017}. Though preliminary, these results provide a first systematic and robust characterization of cold pools over the wide range of conditions observed during EUREC$^4$A, demonstrating how valuable detecting cold pools from soundings can be for the study of cold pools and cloud-circulation coupling. 

\section{Conclusions}

In this study, we presented a new method using the mixed layer height as a criterion for detecting cold pools from soundings over tropical oceans. This method is based on the analysis of more than 2000 radiosondes and dropsondes from the EUREC$^4$A field campaign that took place over the western tropical Atlantic near Barbados in January-February 2020. The mixed layer height $H_{mix}$, defined using the virtual potential temperature, shows a left heavy tail for EUREC$^4$A soundings that we attribute to cold pools. It leads us to classify the soundings with $H_{mix} < 400 \textrm{ m}$ as cold pool soundings, and those with $H_{mix} \geq 500 \textrm{ m}$ as environmental soundings. 

We first test this criterion in a simulation over the Barbados region, performed with the LES version of the ICON model. In this simulation, cold pools are visually identified as regions with negative surface temperature anomalies and positive surface precipitation. They coincide with regions of shallow mixed layer, supporting the relevance of the mixed layer height as a proxy to identify cold pools in this LES simulation despite the presence of a strong SST gradient across the domain. Then, we apply our detection method to EUREC$^4$A soundings, looking first at January 24 and February 2, two EUREC$^4$A flight days during which the number of cold pools detected in dropsondes is particularly high. On these two days, the cold pools detected with our method are consistent with satellite images, surface time series from the research vessel Meteor, and thermodynamic and dynamic profiles measured by soundings. On February 2, the calculation of the divergence in and above one of the cold pools further reveals intense mesoscale circulations, consistent with the observed persistence of cloud clusters above cold pools that day.

Finally, we give a few statistics for the entire EUREC$^4$A period, focusing initially on the dropsondes launched by HALO in the EUREC$^4$A circle. By comparing, circle by circle, the cold pool and environmental characteristics, we find that at 100 m, the cold pool soundings are on average 1 K colder and 1 g kg$^{-1}$ moister than the environmental soundings at the same altitude, and experience a wind deviation of more than 3 m.s$^{-1}$ with respect to their mean environment. We further show that these differences have a similar amplitude as the vertical differences between 100 and 500 m for each cold pool sounding, while environmental soundings exhibit a small difference between these two altitudes, consistent with the well-mixed nature of the subcloud layer (SCL) in trade-wind regimes. This significant vertical imprint of cold pools in the SCL is then studied for all the EUREC$^4$A soundings. The coldest cold pools turn out to be those with the largest humidity difference between 100 and 500 m. When there is shear in the cloud layer, we find that the spreading of cold pools is favored downshear, suggesting momentum transport by unsaturated downdrafts feeding cold pools.

These first results and the method described here pave the way for more comprehensive analyses of cold pools over tropical oceans. The quantity and quality of the measurements made during EUREC$^4$A is a great opportunity to perform such analyses. In particular, drones and low-flying aircraft are likely to provide valuable information on the vertical structure of cold pools, which could complement sounding measurements. In parallel, time series from research vessels and surface weather stations could help study the characteristics of cold pools close to the ground and better understand their interactions with the surface. The interest in cold pools in tropical oceans is not new \citep{Zipser1969, Leary1979, Houze1981}, but EUREC$^4$A offers an unprecedented dataset to finely characterize the temporal and spatial properties of cold pools and explore unanswered questions about their effect on the organization and propagation of convection. Hopefully, the method and results described in this paper will help us make the best use of this dataset -- and turn this opportunity into reality. 

\acknowledgments
 L.T.P gratefully acknowledges the funding of his PhD by the AMX program of the Ecole Polytechnique. To access GOES-16 visible data, this study benefited from the IPSL Prodiguer-Ciclad facility which is supported by CNRS, UPMC, Labex L-IPSL and funded by the ANR (Grant \#ANR-10-LABX-0018) and by the European FP7 IS-ENES2 project (Grant \#312979). 
 R.V. acknowledges funding from the European Research Council (ERC) under the European Union's Horizon 2020 research and innovation programme (grand agreement No 694768). 


%
%

\datastatement

ICON primary data can be accessed through the "Mistral" super computer of the German Climate Computing Center / Deutsche Klimarechenzentrum (DKRZ). BCO micro-rain radar data are accessible to the broader community, as detailed in \citet{Stevens2016}. Access to primary data is provided here: \url{https://mpimet.mpg.de/en/science/the-atmosphere-in-the-earth-system/working-groups/tropical-cloud-observation/barbadosstation1/instrumentation-and-data}. Radiosonde and dropsonde data are described respectively in \citet{Stephan2020} and \citet{George2020} and available through the AERIS portal (\url{https://eurec4a.aeris-data.fr/}). Free access to Meteor surface meteorological data and Raman lidar data is also provided by AERIS. The MODIS-Terra picture on January 24 is taken from NASA Worldview (\url{https://worldview.earthdata.nasa.gov}).


%






%
%
%

\bibliographystyle{ametsoc2014}
\bibliography{bibliography}

\begin{thebibliography}{49}
\providecommand{\natexlab}[1]{#1}
\providecommand{\url}[1]{\texttt{#1}}
\renewcommand{\UrlFont}{\rmfamily}
\providecommand{\urlprefix}{URL }
\expandafter\ifx\csname urlstyle\endcsname\relax
  \providecommand{\doi}[1]{doi:\discretionary{}{}{}#1}\else
  \providecommand{\doi}{doi:\discretionary{}{}{}\begingroup
  \urlstyle{rm}\Url}\fi
\providecommand{\eprint}[2][]{\url{#2}}

\bibitem[{Bony and Stevens(2019)Bony, and Stevens}]{Bony2019}
Bony, S., and B.~Stevens, 2019: Measuring area-averaged vertical motions with
  dropsondes. \textit{Journal of the Atmospheric Sciences}, \textbf{76~(3)},
  767--783, \doi{https://doi.org/10.1175/JAS-D-18-0141.1}.

\bibitem[{Bony et~al.(2017)}]{Bony2017}
Bony, S., and Coauthors, 2017: {EUREC4A}: A field campaign to elucidate the
  couplings between clouds, convection and circulation. \textit{Surveys in
  Geophysics}, \textbf{38~(6)}, 1529--1568,
  \doi{https://doi.org/10.1007/s10712-017-9428-0}.

\bibitem[{Bretherton and Blossey(2017)Bretherton, and Blossey}]{Bretherton2017}
Bretherton, C., and P.~Blossey, 2017: Understanding mesoscale aggregation of
  shallow cumulus convection using large-eddy simulation. \textit{Journal of
  Advances in Modeling Earth Systems}, \textbf{9~(8)}, 2798--2821,
  \doi{https://doi.org/10.1002/2017MS000981}.

\bibitem[{Bryan and Parker(2010)Bryan, and Parker}]{Bryan2010}
Bryan, G.~H., and M.~D. Parker, 2010: Observations of a squall line and its
  near environment using high-frequency rawinsonde launches during {VORTEX2}.
  \textit{Monthly weather review}, \textbf{138~(11)}, 4076--4097,
  \doi{https://doi.org/10.1175/2010MWR3359.1}.

\bibitem[{Canut et~al.(2012)Canut, Couvreux, Lothon, Pino,, and
  Sa{\"\i}d}]{Canut2012}
Canut, G., F.~Couvreux, M.~Lothon, D.~Pino, and F.~Sa{\"\i}d, 2012:
  Observations and large-eddy simulations of entrainment in the sheared
  {Sahelian} boundary layer. \textit{Boundary-layer meteorology},
  \textbf{142~(1)}, 79--101, \doi{https://doi.org/10.1007/s10546-011-9661-x}.

\bibitem[{Craig~Goff(1976)}]{Craig1976}
Craig~Goff, R., 1976: Vertical structure of thunderstorm outflows.
  \textit{Monthly Weather Review}, \textbf{104~(11)}, 1429--1440,
  \doi{https://doi.org/10.1175/1520-0493(1976)104<1429:VSOTO>2.0.CO;2}.

\bibitem[{Cuijpers and Duynkerke(1993)Cuijpers, and Duynkerke}]{Cuijpers1993}
Cuijpers, J., and P.~Duynkerke, 1993: Large eddy simulation of trade wind
  cumulus clouds. \textit{Journal of the Atmospheric Sciences},
  \textbf{50~(23)}, 3894--3908,
  \doi{https://doi.org/10.1175/1520-0469(1993)050<3894:LESOTW>2.0.CO;2}.

\bibitem[{de~Szoeke et~al.(2017)de~Szoeke, Skyllingstad, Zuidema,, and
  Chandra}]{Szoeke2017}
de~Szoeke, S.~P., E.~D. Skyllingstad, P.~Zuidema, and A.~S. Chandra, 2017: Cold
  pools and their influence on the tropical marine boundary layer.
  \textit{Journal of the Atmospheric Sciences}, \textbf{74~(4)}, 1149--1168,
  \doi{https://doi.org/10.1175/JAS-D-16-0264.1}.

\bibitem[{Dipankar et~al.(2015)Dipankar, Stevens, Heinze, Moseley, Z{\"a}ngl,
  Giorgetta,, and Brdar}]{Dipankar2015}
Dipankar, A., B.~Stevens, R.~Heinze, C.~Moseley, G.~Z{\"a}ngl, M.~Giorgetta,
  and S.~Brdar, 2015: Large eddy simulation using the general circulation model
  {ICON}. \textit{Journal of Advances in Modeling Earth Systems},
  \textbf{7~(3)}, 963--986, \doi{https://doi.org/10.1002/2015MS000431}.

\bibitem[{Drager and van~den Heever(2017)Drager, and van~den
  Heever}]{Drager2017}
Drager, A.~J., and S.~C. van~den Heever, 2017: Characterizing convective cold
  pools. \textit{Journal of Advances in Modeling Earth Systems},
  \textbf{9~(2)}, 1091--1115, \doi{https://doi.org/10.1002/2016MS000788}.

\bibitem[{Droegemeier and Wilhelmson(1985)Droegemeier, and
  Wilhelmson}]{Droegemeier1985}
Droegemeier, K.~K., and R.~B. Wilhelmson, 1985: Three-dimensional numerical
  modeling of convection produced by interacting thunderstorm outflows. {Part
  II}: Variations in vertical wind shear. \textit{Journal of the atmospheric
  sciences}, \textbf{42~(22)}, 2404--2414,
  \doi{https://doi.org/10.1175/1520-0469(1985)042<2404:TDNMOC>2.0.CO;2}.

\bibitem[{Feng et~al.(2015)Feng, Hagos, Rowe, Burleyson, Martini,, and
  de~Szoeke}]{Feng2015}
Feng, Z., S.~Hagos, A.~K. Rowe, C.~D. Burleyson, M.~N. Martini, and S.~P.
  de~Szoeke, 2015: Mechanisms of convective cloud organization by cold pools
  over tropical warm ocean during the amie/dynamo field campaign.
  \textit{Journal of Advances in Modeling Earth Systems}, \textbf{7~(2)},
  357--381, \doi{https://doi.org/10.1002/2014MS000384},
  \urlprefix\url{https://agupubs.onlinelibrary.wiley.com/doi/abs/10.1002/2014MS000384},
  \eprint{https://agupubs.onlinelibrary.wiley.com/doi/pdf/10.1002/2014MS000384}.

\bibitem[{George~et al.(2020)}]{George2020}
George~et al., G., 2020: {Dropsondes during EUREC4A}, manuscript submitted for
  publication.

\bibitem[{Grandpeix and Lafore(2010)Grandpeix, and Lafore}]{Grandpeix2010}
Grandpeix, J.-Y., and J.-P. Lafore, 2010: A density current parameterization
  coupled with {Emanuel’s} convection scheme. {Part I}: The models.
  \textit{Journal of the Atmospheric Sciences}, \textbf{67~(4)}, 881--897,
  \doi{https://doi.org/10.1175/2009JAS3044.1}.

\bibitem[{Grant et~al.(2020)Grant, Moncrieff, Lane,, and van~den
  Heever}]{Grant2020}
Grant, L.~D., M.~W. Moncrieff, T.~P. Lane, and S.~C. van~den Heever, 2020:
  Shear-parallel tropical convective systems: Importance of cold pools and wind
  shear. \textit{Geophysical Research Letters}, \textbf{47~(12)},
  e2020GL087\,720, \doi{https://doi.org/10.1029/2020GL087720}.

\bibitem[{Houze~Jr and Betts(1981)Houze~Jr, and Betts}]{Houze1981}
Houze~Jr, R.~A., and A.~K. Betts, 1981: Convection in {GATE}. \textit{Reviews
  of Geophysics}, \textbf{19~(4)}, 541--576,
  \doi{https://doi.org/10.1029/RG019i004p00541}.

\bibitem[{Jeevanjee and Romps(2015)Jeevanjee, and Romps}]{Jeevanjee2015}
Jeevanjee, N., and D.~M. Romps, 2015: Effective buoyancy, inertial pressure,
  and the mechanical generation of boundary layer mass flux by cold pools.
  \textit{Journal of the Atmospheric Sciences}, \textbf{72~(8)}, 3199--3213,
  \doi{https://doi.org/10.1175/JAS-D-14-0349.1}.

\bibitem[{Knupp and Cotton(1982)Knupp, and Cotton}]{Knupp1982}
Knupp, K.~R., and W.~R. Cotton, 1982: An intense, quasi-steady thunderstorm
  over mountainous terrain. {Part II}: Doppler radar observations of the storm
  morphological structure. \textit{Journal of the Atmospheric Sciences},
  \textbf{39~(2)}, 343--358,
  \doi{https://doi.org/10.1175/1520-0469(1982)039<0343:AIQSTO>2.0.CO;2}.

\bibitem[{Kurowski et~al.(2018)Kurowski, Suselj, Grabowski,, and
  Teixeira}]{Kurowski2018}
Kurowski, M.~J., K.~Suselj, W.~W. Grabowski, and J.~Teixeira, 2018:
  Shallow-to-deep transition of continental moist convection: Cold pools,
  surface fluxes, and mesoscale organization. \textit{Journal of the
  Atmospheric Sciences}, \textbf{75~(12)}, 4071--4090,
  \doi{https://doi.org/10.1175/JAS-D-18-0031.1}.

\bibitem[{Langhans and Romps(2015)Langhans, and Romps}]{Langhans2015}
Langhans, W., and D.~M. Romps, 2015: The origin of water vapor rings in
  tropical oceanic cold pools. \textit{Geophysical Research Letters},
  \textbf{42~(18)}, 7825--7834, \doi{https://doi.org/10.1002/2015GL065623}.

\bibitem[{Leary and Houze~Jr(1979)Leary, and Houze~Jr}]{Leary1979}
Leary, C.~A., and R.~A. Houze~Jr, 1979: The structure and evolution of
  convection in a tropical cloud cluster. \textit{Journal of the Atmospheric
  Sciences}, \textbf{36~(3)}, 437--457,
  \doi{https://doi.org/10.1175/1520-0469(1979)036<0437:TSAEOC>2.0.CO;2}.

\bibitem[{Lenschow and Sun(2007)Lenschow, and Sun}]{Lenschow2007}
Lenschow, D.~H., and J.~Sun, 2007: The spectral composition of fluxes and
  variances over land and sea out to the mesoscale. \textit{Boundary-Layer
  Meteorology}, \textbf{125~(1)}, 63--84,
  \doi{https://doi.org/10.1007/s10546-007-9191-8}.

\bibitem[{Mahoney et~al.(2009)Mahoney, Lackmann,, and Parker}]{Mahoney2009}
Mahoney, K.~M., G.~M. Lackmann, and M.~D. Parker, 2009: The role of momentum
  transport in the motion of a quasi-idealized mesoscale convective system.
  \textit{Monthly weather review}, \textbf{137~(10)}, 3316--3338,
  \doi{https://doi.org/10.1175/2009MWR2895.1}.

\bibitem[{Medeiros and Nuijens(2016)Medeiros, and Nuijens}]{Medeiros2016}
Medeiros, B., and L.~Nuijens, 2016: Clouds at {Barbados} are representative of
  clouds across the trade wind regions in observations and climate models.
  \textit{Proceedings of the National Academy of Sciences}, \textbf{113~(22)},
  E3062--E3070, \doi{https://doi.org/10.1073/pnas.1521494113}.

\bibitem[{Meyer and Haerter(2020)Meyer, and Haerter}]{Meyer2020}
Meyer, B., and J.~O. Haerter, 2020: Mechanical forcing of convection by cold
  pools: Collisions and energy scaling. \textit{Journal of Advances in Modeling
  Earth Systems}, \textbf{12~(11)}, e2020MS002\,281,
  \doi{https://doi.org/10.1029/2020MS002281},
  \urlprefix\url{https://agupubs.onlinelibrary.wiley.com/doi/abs/10.1029/2020MS002281},
  e2020MS002281 10.1029/2020MS002281,
  \eprint{https://agupubs.onlinelibrary.wiley.com/doi/pdf/10.1029/2020MS002281}.

\bibitem[{Moncrieff(1992)}]{Moncrieff1992}
Moncrieff, M.~W., 1992: Organized convective systems: Archetypal dynamical
  models, mass and momentum flux theory, and parametrization. \textit{Quarterly
  Journal of the Royal Meteorological Society}, \textbf{118~(507)}, 819--850,
  \doi{https://doi.org/10.1002/qj.49711850703}.

\bibitem[{Pennell and LeMone(1974)Pennell, and LeMone}]{Pennell1974}
Pennell, W., and M.~LeMone, 1974: An experimental study of turbulence structure
  in the fair-weather trade wind boundary layer. \textit{Journal of the
  Atmospheric Sciences}, \textbf{31~(5)}, 1308--1323,
  \doi{https://doi.org/10.1175/1520-0469(1974)031<1308:AESOTS>2.0.CO;2}.

\bibitem[{Purdom(1976)}]{Purdom1976}
Purdom, J.~F., 1976: Some uses of high-resolution {GOES} imagery in the
  mesoscale forecasting of convection and its behavior. \textit{Monthly Weather
  Review}, \textbf{104~(12)}, 1474--1483,
  \doi{https://doi.org/10.1175/1520-0493(1976)104<1474:SUOHRG>2.0.CO;2}.

\bibitem[{Rauber et~al.(2007)}]{Rauber2007}
Rauber, R.~M., and Coauthors, 2007: {Rain in Shallow Cumulus Over the Ocean:
  The RICO Campaign}. \textit{Bulletin of the American Meteorological Society},
  \textbf{88~(12)}, 1912--1928, \doi{https://doi.org/10.1175/BAMS-88-12-1912}.

\bibitem[{Rochetin et~al.(2020)Rochetin, Hohenegger, Touz{\'e}-Peiffer,, and
  Villefranque}]{Rochetin2020}
Rochetin, N., C.~Hohenegger, L.~Touz{\'e}-Peiffer, and N.~Villefranque, 2020: A
  physically-based robust definition of convectively generated density
  currents: detection and characterization in convection-permitting
  simulations, manuscript submitted for publication.

\bibitem[{Rotunno et~al.(1988)Rotunno, Klemp,, and Weisman}]{Rotunno1988}
Rotunno, R., J.~B. Klemp, and M.~L. Weisman, 1988: A theory for strong,
  long-lived squall lines. \textit{Journal of the Atmospheric Sciences},
  \textbf{45~(3)}, 463--485,
  \doi{https://doi.org/10.1175/1520-0469(1988)045<0463:ATFSLL>2.0.CO;2}.

\bibitem[{Schlemmer and Hohenegger(2014)Schlemmer, and
  Hohenegger}]{Schlemmer2014}
Schlemmer, L., and C.~Hohenegger, 2014: The formation of wider and deeper
  clouds as a result of cold-pool dynamics. \textit{Journal of the Atmospheric
  Sciences}, \textbf{71~(8)}, 2842--2858,
  \doi{https://doi.org/10.1175/JAS-D-13-0170.1}.

\bibitem[{Seifert and Heus(2013)Seifert, and Heus}]{Seifert2013}
Seifert, A., and T.~Heus, 2013: Large-eddy simulation of organized
  precipitating trade wind cumulus clouds. \textit{Atmospheric Chemistry and
  Physics}, \textbf{13}, 5631--5645,
  \doi{https://doi.org/10.5194/acp-13-5631-2013}.

\bibitem[{Stephan et~al.(2020)}]{Stephan2020}
Stephan, C.~C., and Coauthors, 2020: Ship-and island-based atmospheric
  soundings from the 2020 {EUREC4A} field campaign. \textit{Earth System
  Science Data Discussions}, 1--35,
  \doi{https://doi.org/10.5194/essd-2020-174}.

\bibitem[{Stevens et~al.(2016)}]{Stevens2016}
Stevens, B., and Coauthors, 2016: The {Barbados Cloud Observatory}: Anchoring
  investigations of clouds and circulation on the edge of the {ITCZ}.
  \textit{Bulletin of the American Meteorological Society}, \textbf{97~(5)},
  787--801, \doi{https://10.1175/BAMS-D-14-00247.1}.

\bibitem[{Stevens et~al.(2019)}]{Stevens2019}
Stevens, B., and Coauthors, 2019: Sugar, gravel, fish and flowers: Mesoscale
  cloud patterns in the trade winds. \textit{Quarterly Journal of the Royal
  Meteorological Society}, 1--12, \doi{https://doi.org/10.1002/qj.3662}.

\bibitem[{Stevens~et al.(2020)}]{Stevens2020overview}
Stevens~et al., B., 2020: {EUREC4A}, manuscript submitted for publication.

\bibitem[{Tompkins(2001)}]{Tompkins2001}
Tompkins, A.~M., 2001: Organization of tropical convection in low vertical wind
  shears: The role of cold pools. \textit{Journal of the Atmospheric Sciences},
  \textbf{58~(13)}, 1650--1672,
  \doi{https://doi.org/10.1175/1520-0469(2001)058<1650:OOTCIL>2.0.CO;2}.

\bibitem[{Vial et~al.(2019)}]{Vial2019}
Vial, J., and Coauthors, 2019: A new look at the daily cycle of trade wind
  cumuli. \textit{Journal of advances in modeling earth systems},
  \textbf{11~(10)}, 3148--3166, \doi{https://doi.org/10.1029/2019MS001746}.

\bibitem[{Vogel(2017)}]{Vogel2017}
Vogel, R., 2017: The influence of precipitation and convective organization on
  the structure of the trades. Ph.D. thesis, Universit{\"a}t Hamburg Hamburg,
  \doi{https://doi.org/10.17617/2.2503092}.

\bibitem[{Warner et~al.(1979)Warner, Simpson, Martin, Suchman, Mosher,, and
  Reinking}]{Warner1979}
Warner, C., J.~Simpson, D.~Martin, D.~Suchman, F.~Mosher, and R.~Reinking,
  1979: Shallow convection on day 261 of {GATE}/mesoscale arcs. \textit{Monthly
  Weather Review}, \textbf{107~(12)}, 1617--1635,
  \doi{https://doi.org/10.1175/1520-0493(1979)107<1617:SCODOG>2.0.CO;2}.

\bibitem[{Webster and Lukas(1992)Webster, and Lukas}]{Webster1992}
Webster, P.~J., and R.~Lukas, 1992: {TOGA COARE}: The coupled ocean--atmosphere
  response experiment. \textit{Bulletin of the American Meteorological
  Society}, \textbf{73~(9)}, 1377--1416,
  \doi{https://doi.org/10.1175/1520-0477(1992)073<1377:TCTCOR>2.0.CO;2}.

\bibitem[{Weisman and Rotunno(2004)Weisman, and Rotunno}]{Weisman2004}
Weisman, M.~L., and R.~Rotunno, 2004: “a theory for strong long-lived squall
  lines” revisited. \textit{Journal of the Atmospheric Sciences},
  \textbf{61~(4)}, 361--382,
  \doi{https://doi.org/10.1175/1520-0469(2004)061<0361:ATFSLS>2.0.CO;2}.

\bibitem[{Yoneyama et~al.(2013)Yoneyama, Zhang,, and Long}]{Yoneyama2013}
Yoneyama, K., C.~Zhang, and C.~N. Long, 2013: Tracking pulses of the
  {Madden--Julian} oscillation. \textit{Bulletin of the American Meteorological
  Society}, \textbf{94~(12)}, 1871--1891,
  \doi{https://doi.org/10.1175/BAMS-D-12-00157.1}.

\bibitem[{Young et~al.(1995)Young, Perugini,, and Fairall}]{Young1995}
Young, G.~S., S.~M. Perugini, and C.~W. Fairall, 1995: Convective wakes in the
  equatorial western {Pacific} during {TOGA}. \textit{Monthly Weather Review},
  \textbf{123~(1)}, 110--123,
  \doi{https://doi.org/10.1175/1520-0493(1995)123<0110:CWITEW>2.0.CO;2}.

\bibitem[{Z{\"a}ngl et~al.(2015)Z{\"a}ngl, Reinert, R{\'\i}podas,, and
  Baldauf}]{Zangl2015}
Z{\"a}ngl, G., D.~Reinert, P.~R{\'\i}podas, and M.~Baldauf, 2015: The {ICON
  (ICOsahedral Non-hydrostatic)} modelling framework of {DWD} and {MPI-M}:
  Description of the non-hydrostatic dynamical core. \textit{Quarterly Journal
  of the Royal Meteorological Society}, \textbf{141~(687)}, 563--579,
  \doi{https://doi.org/10.1002/qj.2378}.

\bibitem[{Zipser(1969)}]{Zipser1969}
Zipser, E.~J., 1969: The role of organized unsaturated convective downdrafts in
  the structure and rapid decay of an equatorial disturbance. \textit{Journal
  of Applied Meteorology}, \textbf{8~(5)}, 799--814,
  \doi{https://doi.org/10.1175/1520-0450(1969)008<0799:TROOUC>2.0.CO;2}.

\bibitem[{Zuidema et~al.(2017)Zuidema, Torri, Muller,, and
  Chandra}]{Zuidema2017}
Zuidema, P., G.~Torri, C.~Muller, and A.~Chandra, 2017: A survey of
  precipitation-induced atmospheric cold pools over oceans and their
  interactions with the larger-scale environment. \textit{Surveys in
  Geophysics}, \textbf{38~(6)}, 1283--1305,
  \doi{https://doi.org/10.1007/s10712-017-9447-x}.

\bibitem[{Zuidema et~al.(2012)}]{Zuidema2012}
Zuidema, P., and Coauthors, 2012: On trade wind cumulus cold pools.
  \textit{Journal of the Atmospheric Sciences}, \textbf{69~(1)}, 258--280,
  \doi{https://doi.org/10.1175/JAS-D-11-0143.1}.

\end{thebibliography}

%

%

\end{document}